\pdfoutput=1
\documentclass[a4paper,fleqn,usenatbib]{mnras}

\usepackage{newtxtext,newtxmath}

\usepackage[T1]{fontenc}

\usepackage{graphicx}	
\usepackage{amsmath}	
\usepackage{amssymb}	
\usepackage{bm}
\usepackage{ulem}
\usepackage{xspace}
\usepackage{threeparttable}
\usepackage{lineno}

\newcommand\figr{Fig.~}
\newcommand\tabr{Table~}
\newcommand\secr{Section~}
\newcommand\tycho{Tycho's SNR\xspace}
\newcommand\chandra{{\it Chandra}\xspace}
\newcommand\suzaku{{\it Suzaku}\xspace}
\newcommand\Ka{\hbox{K$\alpha$}}
\newcommand\Kb{\hbox{K$\beta$}}
\newcommand\Lya{\hbox{Ly$\alpha$}}
\newcommand\Lyb{\hbox{Ly$\beta$}}
\newcommand\Hea{\hbox{He$\alpha$}}
\newcommand\Heb{\hbox{He$\beta$}}
\newcommand\Heg{\hbox{He$\gamma$}}

\title[X-ray study of Tycho's SNR using deep learning]{X-ray Study of Spatial Structures in Tycho's Supernova Remnant Using Unsupervised Deep Learning}

\author[H. Iwasaki et al.]{
Hiroyoshi Iwasaki,$^{1,2}$\thanks{E-mail: h.iwasaki@rikkyo.ac.jp (HI)}
Yuto Ichinohe,$^{1,2}$
and Yasunobu Uchiyama,$^{1,2}$
\\
$^{1}$Department of Physics, Rikkyo University, 3-34-1 Nishi Ikebukuro, Toshima-ku, Tokyo 171-8501, Japan\\
$^{2}$Bluish AI Laboratory, 7-11-3 Ginza, Chuo-ku, Tokyo 104-0061, Japan\\
}

\date{Accepted XXX. Received YYY; in original form ZZZ}

\pubyear{2019}

\begin{document}
\label{firstpage}
\pagerange{\pageref{firstpage}--\pageref{lastpage}}
\maketitle

\begin{abstract}
Recent rapid development of deep learning algorithms, which can implicitly capture structures in high-dimensional data, opens a new chapter in astronomical data analysis. We report here a new implementation of deep learning techniques for X-ray analysis. We apply a variational autoencoder (VAE) using a deep neural network for spatio-spectral analysis of data obtained by {\it Chandra} X-ray Observatory from Tycho's supernova remnant (SNR). We established an unsupervised learning method combining the VAE and a Gaussian mixture model (GMM), where the dimensions of the observed spectral data are reduced by the VAE, and clustering in feature space is performed by the GMM. We found that some characteristic spatial structures, such as the iron knot on the eastern rim, can be automatically recognised by this method, which uses only spectral properties. This result shows that unsupervised machine learning can be useful for extracting characteristic spatial structures from spectral information in observational data (without detailed spectral analysis), which would reduce human-intensive preprocessing costs for understanding fine structures in diffuse astronomical objects, e.g., SNRs or clusters of galaxies. Such data-driven analysis can be used to select regions from which to extract spectra for detailed analysis and help us make the best use of the large amount of spectral data available currently and arriving in the coming decades.
\end{abstract}




\section{Introduction}\label{sec:intro}

In the past decade, machine learning, especially deep learning, has occupied an important position in data science because
of its rapid development and high versatility. It has the potential to assist in analysis of astronomical data and to
extract important information from rich astronomical data without human bias.

Astronomical observations produce complex multidimensional data that include spatial, temporal, and spectral
information.
For example, X-ray data obtained from a single observation of a diffuse source, e.g., a supernova remnant (SNR), may contain all these types of information.
Because of this complexity, conventional analyses can be prone to human bias and oversight.
In the near future, a dramatic improvement in the energy resolution of X-ray observations is expected; e.g., {\it XRISM} will have an energy resolution of several electron volts \citep{xrism2018}, and {\it Athena} will have a spectral resolution of 2.5~eV up to 7~keV at a spatial resolution of $\sim$5~arcsec with $\sim$4000~pixels \citep{AthenaX-IFU2018SPIE}.
Detailed analysis of such data may require excessive human resources.
Thus, automatic and unbiased methods to discover features and pre-analyse the data are required to exploit the full potential of upcoming instruments.

To take advantage of the rich data contained in SNR observations and to extract essential information without human bias, various machine learning techniques have been explored.
\citet{Warren2005} and \citet{Warren2006PhDT} demonstrated the separation of mostly featureless and line-dominated emission from \tycho\ using a linear dimensionality reduction method, principal component analysis (PCA).
They extracted 12~new axes from 12~broad spectral channels and found that the image of the first principal axis (PC1) corresponds to the contrast between Si- and Fe-rich emission and the hard continuum emission.
\citet{Sato2017} also performed PCA of the narrow-band spectra of \tycho, separating the Si~\Hea\ band into 18~bins, and found that the first three PCs correspond to the line equivalent width, line energy centroid, and line energy width, respectively.
\citet{Burkey2013} demonstrated clustering of four-band line fluxes extracted from 5000~spatial regions of Kepler's SNR using a Gaussian mixture model (GMM) and identified the shocked circumstellar medium (CSM) region.

Most previous applications of machine learning techniques to analysis of SNR data have been limited to linear methods. However, the value of each spectral bin depends nonlinearly on the underlying physical parameters; e.g., the bremsstrahlung continuum emission is exponentially affected by the plasma temperature. In other words, the data space of X-ray spectra is not flat. Although PCA, which linearly transforms the data into another orthogonal expression, might provide approximate results in some cases, it is reasonable to choose a model capable of expressing nonlinear relations when the problem exhibits such features. Deep neural networks (DNNs) are likely to obtain more effective expressions from data spaces than linear methods because of their ability to handle nonlinear relations. In this research, we examine the potential of DNNs to extract features embedded in nonlinear relations.

An artificial neural network (ANN), which consists of several layers of multiple formal neurons, mimics the functioning of animal brains \citep[e.g.,][]{Lecun2015Nature} and can implicitly capture features embedded in high-dimensional data.
Each neuron transforms its input $\bm x$ as $f(\sum \mathbf W{\bm x}+{\bm b})$,
where $\mathbf{W}$ and $\bm b$ are tuneable weight parameters, and $f({\bm x})$ is a nonlinear function, and the processing layers learn representations of data with multiple levels of abstraction.
Multilayer ANN architectures, such as the multilayer perceptron, can reveal complex, nonlinear relations.

Recent computational advances have made it possible to train DNNs at a reasonable time and cost, and such techniques have become very popular in many areas, including analyses of astrophysical images \citep{Hezaveh2017Nature, Kimura2017arXiv},
spectra \citep{Ichinohe2018MNRAS, Ichinohe2019}, light curves \citep{Charnock2017ApJ, Shallue2018AJ},
and telescope events \citep{Shilon2018arXiv}.

This paper is organised as follows.
In Section~\ref{sec:ml_methods}, we describe our machine learning method.
\secr\ref{sec:demo} presents our machine learning results and interpretation of the features classified by the method. In \secr\ref{sec:analysis}, we analyse the two regions selected by our method in detail.
Section~\ref{sec:discussion} discusses the method's efficiency and concludes our paper.

\section{Machine Learning Methods}\label{sec:ml_methods}

\begin{figure}
\centering
\includegraphics[width=\columnwidth,keepaspectratio]{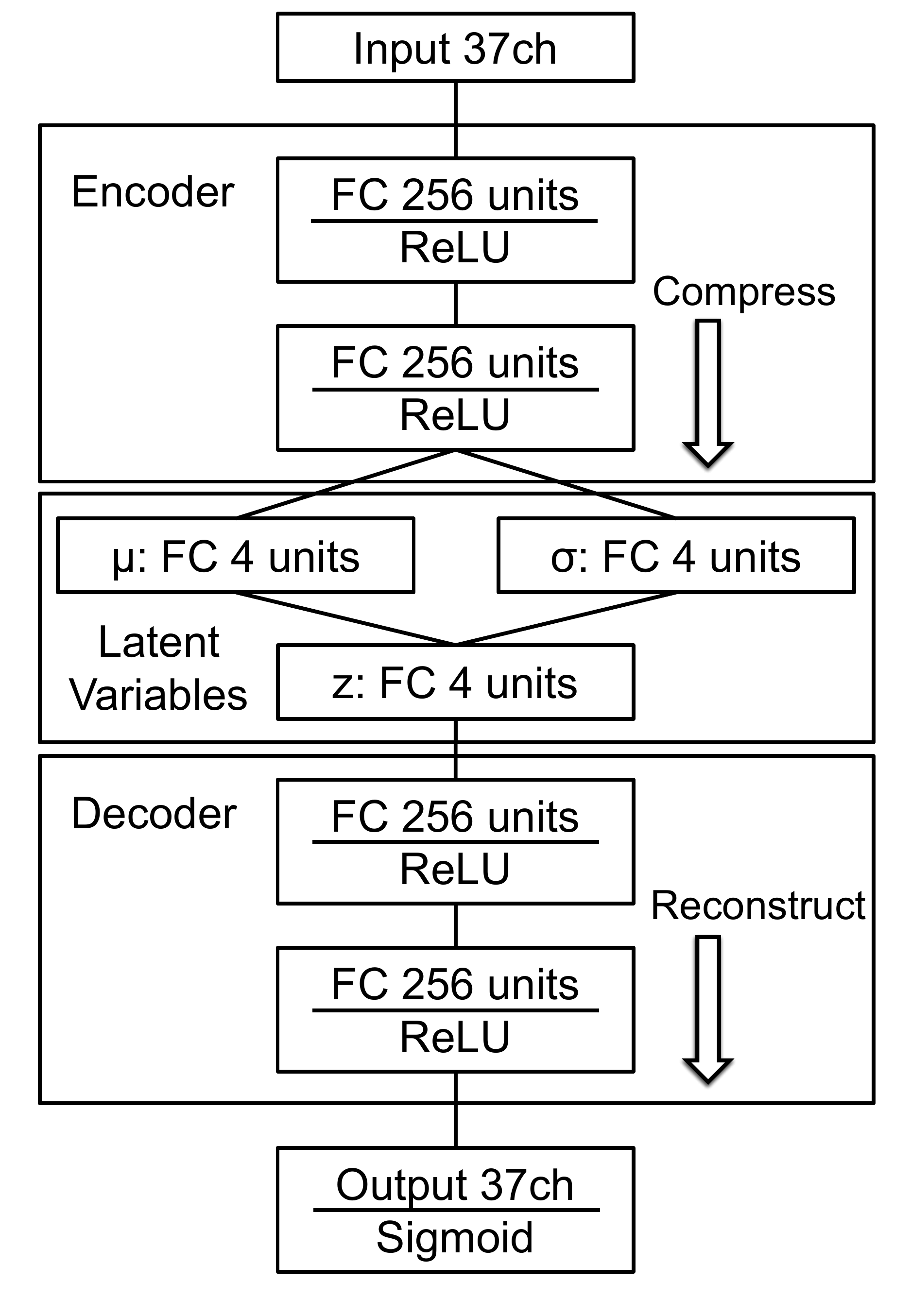}
\caption{Diagram of the VAE architecture. See the main text for the details of the network.}
\label{fig:VAE_architecture}
\end{figure}

In this work, we aim to categorise individual spatial bins of an X-ray image using only the spectral properties.
We developed a method that allows us to combine two unsupervised learning methods, the
variational autoencoder \citep[VAE;][]{Kingma2013} and the GMM, for spatially resolved spectroscopy of the X-ray data obtained by \chandra\ from \tycho.

Similar methods in which dimensionality reduction and unsupervised clustering are integrated have already been applied in several astronomical analyses,
e.g., classification of galaxy spectra \citep[PCA \& GMM;][]{Hurley2012MNRAS},
classification of supernova \citep[autoencoder (AE), isomap \& k-means;][]{Ishida2017IAUS} or separation of galactic and extragalactic objects \citep[AE \& support vector machine (SVM);][]{Khramtsov2018arXiv}.
In this paper, we explore the method of combining nonlinear dimensionality reduction by a VAE and clustering by a GMM for automatic investigation of the spatial structures of a diffuse object for the first time.

\subsection{Variational Autoencoder}\label{sec:vae}

Although an X-ray spectrum of, e.g., an SNR, may have a large number of energy bins, they are not completely independent of each other because of the finite instrumental energy resolution and underlying emission process.
The number of essential parameters generating the spectrum might be small (e.g., APEC, an emission model for plasma in collisional ionization equilibrium, has only four parameters).
Therefore, before classifying the spectrum, we reduced the dimensionality by compressing the information in the raw input spectra.
To capture nonlinear relationships between the essential parameters, we employed an unsupervised DNN architecture, namely, a VAE.

An AE is a DNN architecture connecting an encoder and a decoder.
The encoder is trained to encode the input data as latent variables by reducing the input dimensions.
At the same time, the decoder is trained to reproduce the original input from the latent variables $\bm z$.
The dimension of the latent space is smaller than that of the original space.
One can thus obtain a compressed latent expression of the original data if the AE is successfully trained.
The latent space is expected to capture nonlinear relationships in the input data because of the capabilities of DNNs.

The VAE is a variant of the AE. In the VAE, a multidimensional Gaussian distribution is assumed for the latent variables.
Unlike a normal AE, which computes
the latent variables directly, the encoder of a VAE computes the means $\bm\mu$ and variances $\bm\sigma$.
A set of latent variables $\bm z$ is sampled from a multidimensional Gaussian whose means and variances are calculated by the encoder.
The decoder decompresses the set of latent variables $\bm z$.

VAE models have some advantages over normal AEs, e.g., more stable training and a better latent manifold structure \citep{Tolstikhin2017arXiv}.
Before the study, we compared normal AEs to the VAE and found that training of the normal AE is sometimes unstable and that in many cases, some latent variables do not represent any features.
Therefore, we chose the VAE for this study.

Figure \ref{fig:VAE_architecture} shows a diagram of the VAE architecture.
The encoder and decoder were constructed from two fully connected (FC) neural network layers with 256~nodes per layer and rectified linear unit \citep[ReLU;][]{nair2010relu} activation.

The encoder branches after the FC layers and connects to a layer of $\bm\mu$ and $\bm\sigma$, which have the same number of nodes as $\bm z$.
The latent variables $\bm z$ are sampled as $\bm z = \bm\mu + \bm\sigma \odot \bm\epsilon$\footnote{Here, $\odot$ is the operator of the element-wise product.} using the {\it reparametrization trick} \citep{Kingma2013, Rezende2014}.
Here $\bm\epsilon$ is a vector consisting of random numbers sampled from a Gaussian distribution defined by hyperparameters; the mean is 0, and the variance is 1.
The decoder consists of two layers; the output layer has sigmoid activation and the same dimensions as the input layer.

We used the deep learning framework Keras~2.0.7 \citep{chollet2015keras} with the Tensorflow~1.3.0 \citep{tensorflow2015-whitepaper} backend.
The sum of the binary cross-entropy and KL divergence was used as the loss function.
Nesterov-accelerated adaptive moment estimation \citep[Nadam;][]{Dozat2016} (which, according to our tests, provides the fastest convergence of the optimisers) was used for optimisation.
The training was performed for 100~epochs with a batch size of 100.

Training typically takes 16--20~min on a workstation with an eight-core Intel Xeon E5 CPU.
We also tested the training of a model using GPU computations on an NVIDIA GeForce GTX 1080Ti graphics card. Training on the GPU with a batch size of 4096 typically ran for 40~s and required approximately 300~MiB of GPU memory.

\subsection{Gaussian Mixture Model}\label{sec:gmm}

After the spectral information was compressed into a several-dimensional latent expression using the VAE as explained in the previous section, we classified it using the GMM.

There are two major types of clustering methods: (1) hard clustering methods, in which each data point is assigned to
only one category (e.g., k-means, SVM), and (2) soft clustering methods, which assign each point to all the
categories with different weights (e.g., the GMM).

A soft clustering method is appropriate for this study for the following reasons. (1) In the latent variable coded by the VAE, the characteristic distribution is not necessarily separate, and components with apparently different trends (see \figr\ref{fig:GMM_Tycho_scatter}) overlap considerably, particularly around the value 0. It is generally difficult to draw clear boundaries when multiple components overlap. (2) The physical conditions change continuously throughout the SNR.

The GMM is a well-known soft clustering method that has become popular for astrophysical data analysis \citep[e.g.,][]{Davoodi2006AJ, Hurley2012MNRAS, Burkey2013}.
It describes the data distribution as multidimensional Gaussians; each Gaussian represents a clustering category.
Every data point is represented by a weighted superposition of all the categories.
The probability that a data point belongs to a certain category, which is also referred to as the {\it responsibility}, is represented by the ratio of the value of the Gaussian corresponding to the category to the sum of the values of all the Gaussians for this data point. We used the GMM {GaussianMixture} in scikit-learn~0.19.0 \citep{scikit-learn}, which is a Python library providing a machine learning framework.

\section{Demonstration}\label{sec:demo}

\subsection{Demonstration Target}\label{sec:target}

X-ray spectra from SNRs are known to contain rich multidimensional information. We chose Tycho's SNR as a target for application of our machine learning method because this is one of the best studied SNRs.

Supernovae (SNe) explosively eject elements synthesised in the progenitor materials (so-called ejecta), forming blast waves.
The resulting bright X-ray-emitting structures are called SNRs. X-ray observations of SNRs allow us to investigate both the chemical evolution of the Universe and the mechanisms of cosmic-ray acceleration.

SNe have long been assumed to supply heavy elements synthesised during the explosion. The ejecta, that is, the X-ray-emitting hot plasma in SNRs, reveals the nuclear burning regimes of elemental synthesis.

\tycho\ is the remnant of SN~1572, which is known to be a type Ia explosion from the light-echo spectrum \citep{Krause2008}.
In X-ray spectra of \tycho, line emission from intermediate-mass elements (IMEs; e.g., Si, S, Ar, and Ca) and Fe synthesised during the supernova explosion are clearly seen.
In addition, secondary Fe-peak elements (e.g., Cr, Mn, and Ni, which are synthesised together with Fe) have been detected \citep[e.g.,][]{Tamagawa2009}.
The global morphology of \tycho\ features radial gradation of the plasma ionization state and the electron or ion temperature, which are caused by reverse shock (RS) heating.
The gradation features appear as differences in the peak radii of the emission lines and are seen especially clearly in the northwestern (NW) projected ejecta limb.

X-ray imaging using {\it ASCA} showed that the Fe~K emission clearly peaks at a smaller radius than the Fe~L and IME line emission and that the Fe-K-emitting plasma was hotter and less ionized \citep{Hwang1997, Hwang1998}.
\citet{Warren2005} measured the averaged RS radius as 183~arcsec using the Fe~\Ka\ lines from a {\it Chandra} observation.
\citet{Yamaguchi2014} showed electron heating at the RS on the NW limb using the Fe~\Ka\ and \Kb\ lines. They also measured the RS radius as 158~arcsec using the Fe~\Kb\ lines of immediate postshock, low-ionization ejecta, which peak at a smaller radius than Fe~\Ka\ emission from a relatively highly ionized component.

\citet{Lu2015} used {\it Chandra} observations to show a systematic increase in the S-to-Si line flux ratio with increasing radius resulting from RS propagation in the ejecta and reported the elapsed ionization time since the ejecta was shock-heated.
\citet{Sato2017} also found a gradual increase in the line centroids of Fe~\Ka\ beyond the radius and interpreted it as a difference in the elapsed ionization time.

By contrast, the eastern region exhibits an unusual morphological structure called the Fe knot, where several clumps outrun the forward shock (FS).
Detailed analysis of \suzaku\ and \chandra\ data suggests that the Fe knot did not originate in the deep, dense core of the progenitor white dwarf but was instead synthesised under incomplete Si burning or the $\alpha$-rich freeze-out regime \citep{Yamaguchi2017}.

In addition, galactic SNRs are widely believed to supply cosmic rays up to the `knee' energy of the cosmic-ray spectrum at $10^{15}$~eV, accelerating particles to relativistic energies in their blast waves by diffusive shock acceleration. The accelerated electrons emit the nonthermal X-ray synchrotron emission observed from the limbs of young SNRs \citep[e.g.,][]{Koyama1995Nature, eriksen2011}.

The X-ray synchrotron emission from electrons accelerated at the FS was observed from the limb of \tycho\ \citep{Hwang2002ApJ}.
Cosmic-ray proton acceleration at the FS was also reported \citep{Warren2005}.
\citet{eriksen2011} found nonthermal stripes in the projected interior of the remnants and interpreted them as evidence for particle acceleration to the `knee' energy in regions of enhanced magnetic turbulence.

\tycho, which is one of the brightest SNRs in the X-ray band and has various interior structures, is one of the best benchmark objects for testing a new analysis method.
High spatial--spectral-resolution data from \tycho\ were obtained by \chandra.
In this research, we apply our method to the X-ray data from \tycho\ to investigate the morphological structures without human bias by automatic classification of each spatial point based only on the physical features reflected in the spectrum.

\subsection{Data Set}\label{sec:data_set}

\begin{figure}
\centering
\includegraphics[width=\columnwidth,keepaspectratio]{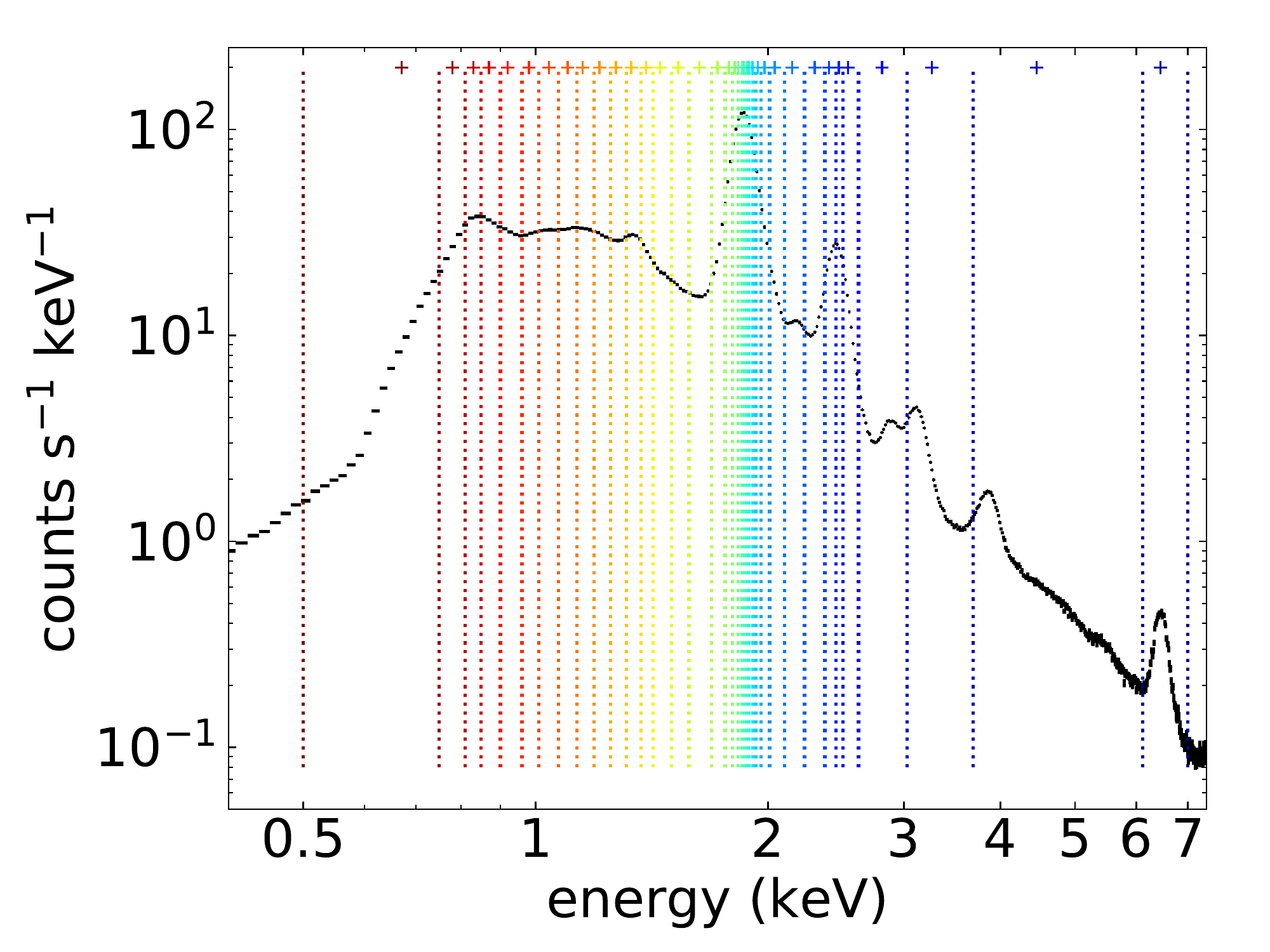}
\caption{Spectrum of entire \tycho. The vertical dotted lines are boundaries of the narrow energy bands. The cross represents the centre of gravity of each energy band.}
\label{fig:Tycho_spec_narrowbands}
\end{figure}

\begin{figure}
\centering
\includegraphics[width=\columnwidth, keepaspectratio]{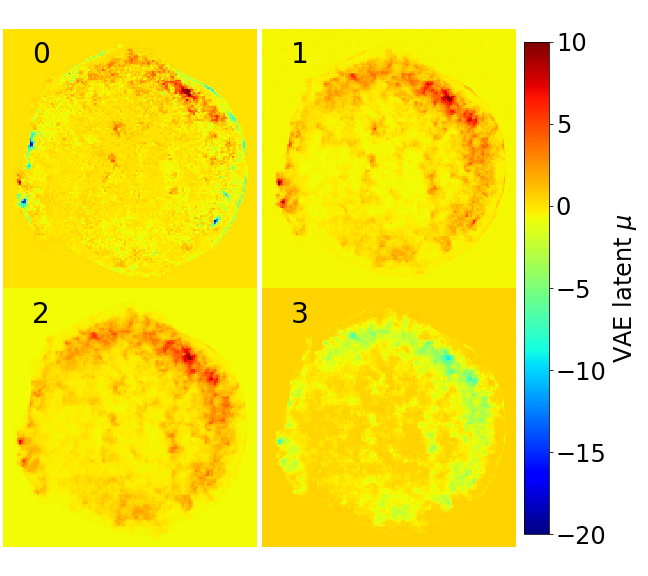}
\caption{Images showing the values of the VAE latent parameters $\bm\mu$ as an image for each axis when the merged data (a 37-colour image observed in 2009) are input. The images, which are standardised to improve the appearance, share a colour scale.
\protect\\
(A colour version of this figure is available in the online journal.)}
\label{fig:VAE_latent_mu_images}
\end{figure}

\begin{figure}
\centering
\includegraphics[width=\columnwidth,keepaspectratio]{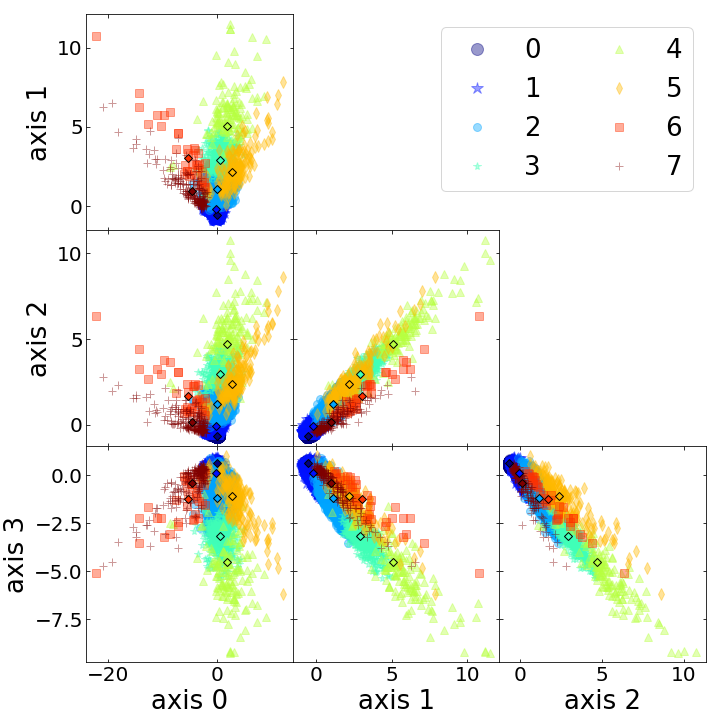}
\caption{Scatter plot showing the $\bm\mu$ of VAE latent parameter obtained by inputting each spatial bin of the merged data observed in 2009, where two of the four axes are chosen.
Each point was colour-coded for each category classified by GMM clustering. The centroid of each category is shown as an open black diamond.
\protect\\
(A colour version of this figure is available in the online journal.)}
\label{fig:GMM_Tycho_scatter}
\end{figure}

\tycho\ was observed by the Advanced CCD Imaging Spectrometer (ACIS)-I of \chandra\ for 145.6, 142.1 (two obsIDs), 734.1 (nine obsIDs), and 146.98~ks in 2003, 2007, 2009, and 2015, respectively.
In 2007 and 2009, there were one and five observations, respectively, with exposure times exceeding 80~ks.
We performed X-ray analysis using CIAO (version 4.9) and CalDB (version 4.7.6) provided by the {\it Chandra} X-ray Center\footnote{Available at \url{http://cxc.harvard.edu}}.

\begin{figure*}
\centering
\includegraphics[width=17cm,keepaspectratio]{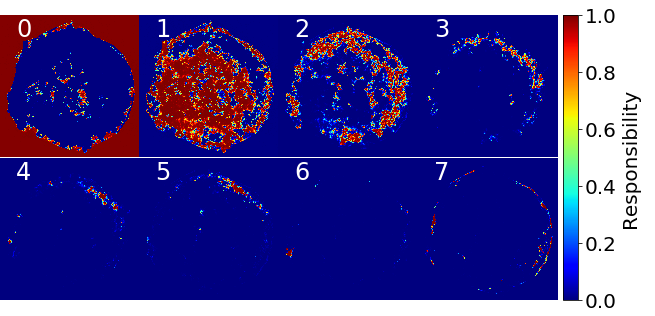}
\caption{GMM responsibility of each spatial bin of merged data set observed in 2009 for each category. The responsibility is between 0 and 1; blue represents a responsibility of 0, and red represents a responsibility of 1.
\protect\\
(A colour version of this figure is available in the online journal.)}
\label{fig:GMM_Tycho_probability}
\end{figure*}

\begin{figure*}
\centering
\includegraphics[height=5.6cm,keepaspectratio]{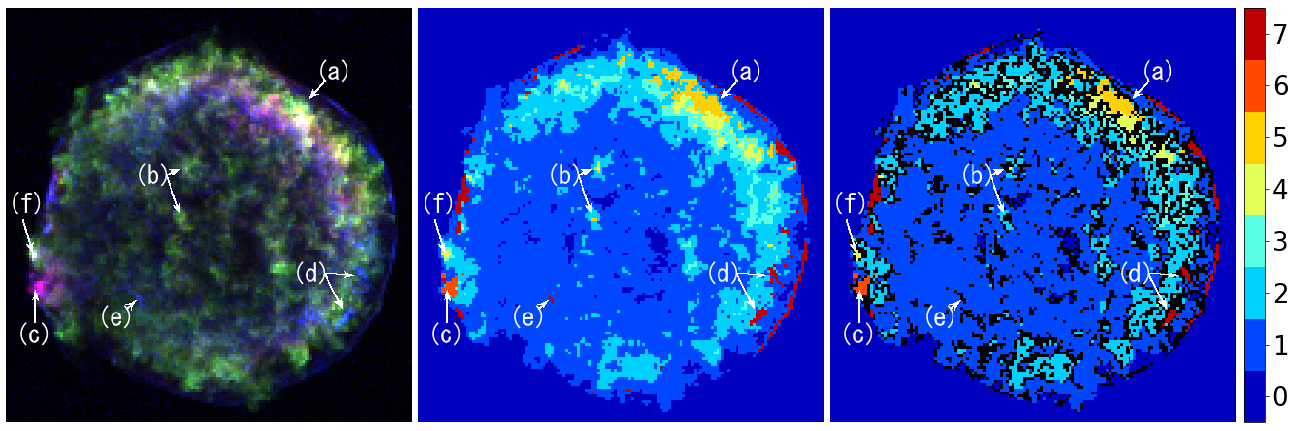}
\caption{ The {\it left panel} shows a three-colour image (red: Fe L blend band, 0.7--0.95~keV; green: Si~\Hea\ band, 1.75--1.95~keV; blue: Fe~\Ka\ band, 6.2--6.9~keV). In the {\it middle panel}, for each spatial bin of the merged data from 2009, the GMM categories with the highest responsibility are assigned and colour-coded. The {\it right panel} shows only spatial bins with a category of more than 90\% responsibility, which are assigned the colours representing the category. The other spatial bins, which have responsibilities below 90\%, appear black.
\protect\\
(A colour version of this figure is available in the online journal.)}
\label{fig:GMM_Tycho_image}
\end{figure*}

Finer spectral binning is expected to preserve more information such as the line width, line-centroid shift, and composition of weak lines. However, finer binning results in lower counts in each spectral bin. We employed an objective method of spectral binning to achieve fine binning and adequate photon statistics in each bin at the same time.
As shown in \figr\ref{fig:Tycho_spec_narrowbands}, the spectrum of the entire \tycho\ was created in the 0.5--7~keV band and was divided into 37 narrower energy bins such that each of them had a count rate of more than 100~$\rm counts\ s^{-1}\ keV^{-1}$, including the background.
The low-energy side (0.5--2.6~keV), with a high count rate, is divided into rather narrow bands (bandwidth/band centroid energy $<$~8\%). For example, the Si~\Hea\ emission line (1.69--2.01~keV) is divided into 11 bins. On the other hand, the high-energy part of the spectrum (2.6--7.0~keV) is divided into four wider bands because the statistics are not as good as at lower energies. The flux values in the 37 narrow energy bands corresponding to a single spatial bin are combined, and the resulting 37-dimensional vectors are used as the input data set.

The flux image of each band was created with the coordinate ranges set to omit regions outside of the SNR. The spatial bin size was set to 3.94~arcsec, resulting in an image size of $146\times 143$~spatial bins. We did not subtract the backgrounds from the images because most of \tycho\ is sufficiently bright that we can safely ignore the contributions from the non-X-ray background and cosmic X-ray background between 0.5 and 7.0~keV.

The averaged expansion velocity of \tycho\ is approximately $\rm 0.3\ arcsec\ yr^{-1}$ \citep{Katsuda2010}, which is significant for the entire data set.
Thus, in our analysis we do not mix observations from different years.
Eight individual observations with exposure times exceeding 80~ks in 2003, 2007, 2009, and 2015 were used for training. In addition, we also used the shorter observation taken in 2007 by co-adding it with the longer one taken in the same year. Eighty~percent of the spatial bins in each flux image were chosen randomly and used as training data, and the rest were used as evaluation data. The actual size of the training and evaluation data sets were 150,781 and 36,808, respectively, excluding the spatial bins with zero flux in all the narrow energy bands (i.e., a 37-dimensional zero vector). All the observations from 2009 (a year which has the longest total exposure) were summed and used for the post-training analysis.

\subsection{Unsupervised Dimension Reduction and Clustering}

We extracted the latent expressions $\bm\mu$ from the data of \tycho\ observed in 2009 using the encoder of the trained VAE.
Each panel in \figr\ref{fig:VAE_latent_mu_images} shows the four-dimensional coordinates of the latent parameters $\bm\mu$ that are obtained by the VAE from a merged set of all the observations from 2009.
The images for each axis of $\bm\mu$ have the same colour scale.
We determined the latent dimension as follows. We prepared models with latent dimensions ranging from 2 to 10 and trained them for 100 epochs. We selected the latent dimension whose corresponding loss value for the validation data was the lowest.

We applied GMM soft clustering to the obtained latent expressions.
The optimal number was determined by checking the Bayesian information criterion for three to nine of the used clusters. There was little difference between cases with seven to nine clusters; thus, for the analysis we used eight clusters.
To visualise the features extracted by the VAE, we plot clusters in different colours in the scatter plot shown in \figr\ref{fig:GMM_Tycho_scatter}.

The scatter plots in \figr\ref{fig:GMM_Tycho_scatter} show the distributions of the latent variables, which are colour-coded according to the category assigned by GMM clustering, projected onto all six (=$_4C_2$, where 4 is the latent dimension) different two-dimensional planes passing through the origin. In the latent space, the compressed expressions form a radial distribution consisting of several branches around 0. The main thick branch is classified as categories~1--4, whereas the three branches extending in different directions are classified as categories~5, 6, and 7, respectively. The data points around 0 are classified as category~0.

Each panel of \figr\ref{fig:GMM_Tycho_probability} shows the responsibility of each GMM category.
The middle panel of \figr\ref{fig:GMM_Tycho_image} shows the division of \tycho\ into GMM categories. The
colours of the spatial bins correspond to the highest responsibility category, as obtained by the method when the merged data from 2009 were used.
The spatial bins of each category have a spatially coherent distribution.

The right panel of \figr\ref{fig:GMM_Tycho_image} shows the same image as the middle panel of \figr\ref{fig:GMM_Tycho_image}, but the spatial bins whose assigned category has a responsibility of $\lid$~90\% are masked. Thus, the spatial bins that remain coloured in the right panel of \figr\ref{fig:GMM_Tycho_image} are robustly assigned to some category, and thus are expected to have some spectral features distinct from those of the other categories.

For comparison, a traditional RGB image is shown in the left panel of \figr\ref{fig:GMM_Tycho_image}. Some regions that appear similar in the RGB image are assigned to different categories. For example, the blob on the eastern rim (region marked `c' in \figr\ref{fig:GMM_Tycho_image}) and the annular layer to the northwest (inner layer region marked `a' in \figr\ref{fig:GMM_Tycho_image}) both appear reddish in the RGB image (\figr\ref{fig:GMM_Tycho_image}, left panel), although they are assigned to different categories in the GMM image (categories 3 and 6; see the middle panel of \figr\ref{fig:GMM_Tycho_image}).

We also note that the clusters corresponding to the layered structure in the NW part of the SNR (for detailed analysis, see \secr\ref{sec:Fe_knot}) are revealed by clustering for any number of Gaussians between seven and nine.
On the other hand, the regions dominated by featureless emission and the Fe knot described in \secr\ref{sec:NW_ejecta} are separated into two clusters only when eight or nine categories are assumed.

\begin{figure}
\centering
\includegraphics[width=\columnwidth,keepaspectratio]{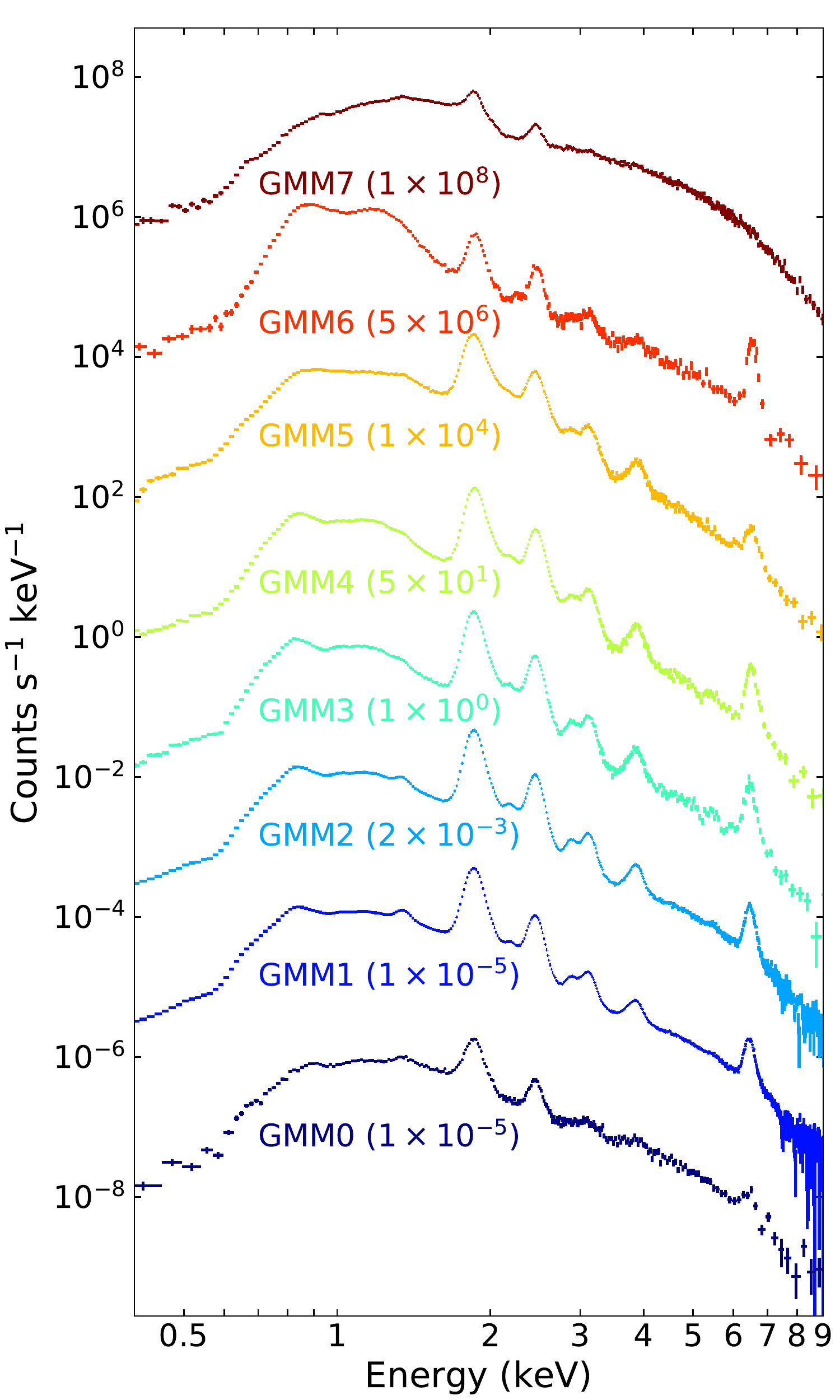}
\caption{Spectrum (background-subtracted) of the region inside the SNR for which each GMM category has a responsibility above 90\%. The normalisation of each spectrum is
 adjusted by multiplying by the factor given in the panel.
\protect\\
(A colour version of this figure is available in the online journal.)}
\label{fig:GMM_Tycho_spec}
\end{figure}

\subsection{Detailed Results of Clustering}\label{sec:spec_ana_GMM_All_tycho}

On the basis of the GMM classification, we extracted the representative spectra of each category by combining all the spatial bins assigned to a certain category with responsibilities above 90\%. The combined spectra are shown in \figr\ref{fig:GMM_Tycho_spec}.
The background was extracted from an annular region surrounding the SNR and subtracted from the spectra.

\begin{figure}
\centering
\includegraphics[height=12cm,keepaspectratio]{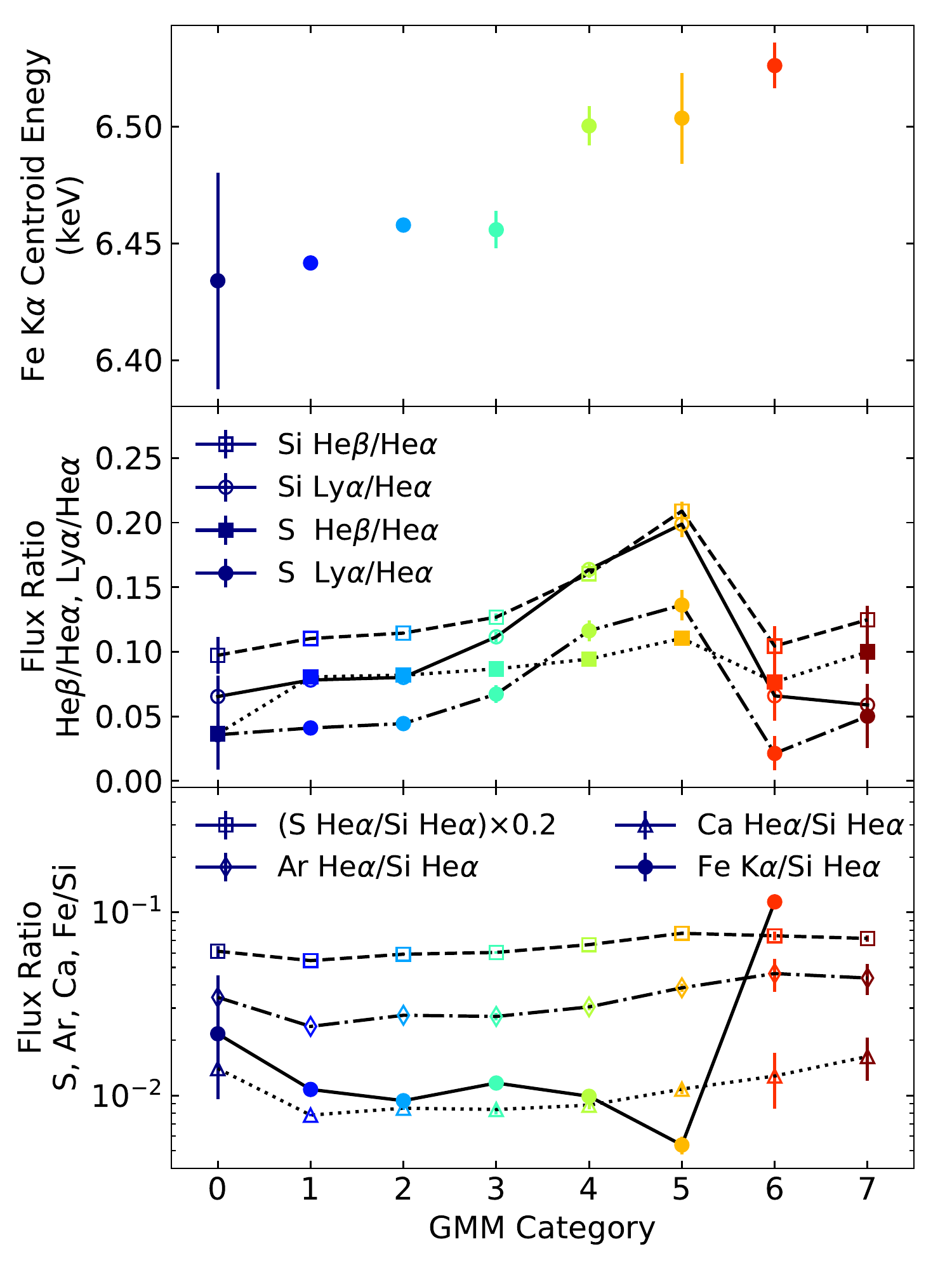}
\caption{Model fitting results of \tycho\ spectra (best-fitting Gaussian parameters of responsibility of each GMM category above 90\%). The {\it top panel} shows the Fe~\Ka\ centroid energies. The {\it middle panel} shows the flux ratios of different transition lines of Si or S (\Lya\ or \Heb\ over \Hea ). The {\it bottom panel} shows the \Ka\ line flux ratios of different elements (S, Ar, Ca, and Fe over Si). The errors are at 90\% C.L.}
\label{fig:GMM_Tycho_line_paras}
\end{figure}

\begin{table}
	\centering
	\caption{GMM categories.}
	\label{tab:GMM_categories}
	\begin{tabular}{cll}
		\hline
		Category No. & Location & Feature \\
		\hline
		0 & outside of SNR & very dark, background \\
		1 & inside of SNR & dark region, between CD and FS \\
		2--4 & rim of ejecta & bright ejecta \\
		5 & NW rim & weak Fe emission\\
		6 & blobs in the Fe knot & strong Fe line emission\\
		7 & rim, filaments & power-law radiation dominant \\
		\hline
	\end{tabular}
\end{table}

\tabr\ref{tab:GMM_categories} summarises the physical interpretation of each category. Category~0 is localised mainly outside of \tycho\ and also contains some dark regions inside the SNR. Categories 1--5, the main regions of which coincide spatially with the ejecta, form a layered structure, and the assigned category numbers, 1--5, change from the inner side of the SNR to the outer side. Category~1 represents faint regions inside of the SNR, which mainly include the unshocked ejecta in projection and the swept-up interstellar medium (ISM)/CSM between the FS and the contact discontinuity (CD). Most of categories~2--5 appears as layered structure, which is most clearly seen in the northern part of the SNR (region marked `a' in \figr\ref{fig:GMM_Tycho_image}),
{especially category~5, which has a responsibility above 90\% and is located only in the NW region of the SNR.}
Two blobs of categories~4 and 5 close to the SNR centre (marked `b' in
\figr\ref{fig:GMM_Tycho_image}) coincide with regions with a measured blue shift \citep{Sato2017}.
Thus, categories~4 and 5 are interpretable as ejecta limbs.

Category~6 is localised at the edge in the eastern part of the SNR (region `c' in \figr\ref{fig:GMM_Tycho_image}), which is associated with the reddish region in the left panel of \figr\ref{fig:GMM_Tycho_image}. This region is a substructure in the Fe knot analysed by \citet{Yamaguchi2017} in detail. The spectrum of category~6 has strong Fe line emission, although the IME emission is weaker in \figr\ref{fig:GMM_Tycho_spec}. As shown in the bottom panel of \figr\ref{fig:GMM_Tycho_line_paras}, category~6 covers a portion of the ejecta having the strongest Fe \Ka\ line in the SNR.

Category~7 corresponds spatially to the FS at the edge of the SNR, the filament and stripe structure at `d' inside the SNR on the western side, and the bright arc at `e' in the SNR on the southeast in \figr\ref{fig:GMM_Tycho_image}. These structures are associated with bluish regions in the left panel of \figr\ref{fig:GMM_Tycho_image}.
In the spectrum of category~7 in \figr\ref{fig:GMM_Tycho_spec}, continuum emission is dominant, although weak contamination by line emission (Si, S, and so on) appears.

\begin{table*}
	\centering
 \caption{Best-fitting parameters for the spectrum of each GMM category with responsibility $\ge$ 0.9}
	\label{tab:GMM_spec_paras}
\begin{threeparttable}
	\begin{tabular}{ccccccccc}
		\hline \hline
		 & \multicolumn{8}{c}{Region} \\
		 & 0\tnote{a} & 1 & 2 & 3 & 4 & 5 & 6 & 7 \\ \hline
		\multicolumn{9}{c}{Fe \Ka\ lines} \\ \hline
		Centroid (keV) &
$6.434^{+0.044}_{-0.046}$ &
$6.442\pm{0.002}$ &
$6.458\pm{0.003}$ &
$6.456\pm{0.008}$ &
$6.500^{+0.007}_{-0.008}$ &
$6.504^{+0.020}_{-0.019}$ &
$6.526^{+0.009}_{-0.010}$ &
-- \\
		Width (keV) &
$0.131^{+0.060}_{-0.054}$ &
$0.088\pm{0.003}$ &
$0.079\pm{0.0043}$ &
$0.094\pm{0.012}$ &
$0.097^{+0.010}_{-0.011}$ &
$0.117\pm{0.032}$ &
$0.060^{+0.014}_{-0.016}$ &
 -- \\
		Normalisation\tnote{b} &
$1.80^{+0.45}_{-0.42}$ &
$317\pm{5}$ &
$126\pm{3}$ &
$15.1\pm{0.8}$ &
$14.8\pm{0.8}$ &
$7.15^{+0.83}_{-0.80}$ &
$6.30\pm{0.45}$ &
-- \\ \hline
		\multicolumn{9}{c}{Line Flux Ratio (\%)} \\ \hline
		Si \Heb /Si \Hea &
$9.7^{+1.5}_{-1.4}$ &
$11.0\pm{0.1}$ &
$11.5\pm{0.1}$ &
$12.7\pm{0.3}$ &
$16.1\pm{0.3}$ &
$20.9^{+0.9}_{-0.7}$ &
$10.4\pm{1.6}$ &
$12.5^{+1.0}_{-1.1}$ \\
		Si \Lya /Si \Hea &
$6.5\pm{1.6}$ &
$7.8\pm{0.1}$ &
$8.0\pm{0.1}$ &
$11.2\pm{0.4}$ &
$16.4\pm{0.4}$ &
$19.9^{+0.8}_{-1.0}$ &
$6.6^{+2.0}_{-1.9}$ &
$5.9\pm{1.0}$ \\
		S \Heb /S \Hea &
$3.7^{+2.4}_{-2.8}$ &
$8.1\pm{0.1}$ &
$8.2\pm{0.1}$ &
$8.7\pm{0.4}$ &
$9.4\pm{0.3}$ &
$11.0\pm{0.5}$ &
$7.7\pm{2.1}$ &
$10.0\pm{1.7}$  \\
		S \Lya /S \Hea &
$3.6^{+5.3}_{-1.9}$ &
$4.1\pm{0.2}$ &
$4.4\pm{0.2}$ &
$6.7\pm{0.7}$ &
$11.6\pm{0.8}$ &
$13.6\pm{1.2}$ &
$2.1^{+2.9}_{-1.3}$ &
$5.0^{+2.3}_{-2.5}$  \\ \hline
		\multicolumn{9}{c}{\Ka\ Line Flux Ratio (\%)} \\ \hline
		S \Hea /Si \Hea &
$30.6^{+1.9}_{-2.1}$ &
$27.2\pm{0.1}$ &
$29.5\pm{0.1}$ &
$30.2\pm{0.4}$ &
$33.3^{+0.4}_{-0.5}$ &
$38.4^{+0.8}_{-0.9}$ &
$37.2\pm{2.2}$ &
$36.0\pm{1.4}$ \\
		Ar \Hea /Si \Hea &
$3.4^{+1.2}_{-1.1}$ &
$2.38^{+0.04}_{-0.05}$ &
$2.74^{+0.04}_{-0.05}$ &
$2.70^{+0.12}_{-0.13}$ &
$3.04\pm{0.13}$ &
$3.87^{+0.19}_{-0.20}$ &
$4.63^{+1.10}_{-0.95}$ &
$4.37^{+0.88}_{-0.83}$ \\
		Ca \Hea /Si \Hea &
$1.40^{+0.52}_{-0.45}$ &
$0.78\pm{0.01}$ &
$0.85\pm{0.02}$ &
$0.84\pm{0.05}$ &
$0.89\pm{0.05}$ &
$1.08\pm{0.06}$ &
$1.28^{+0.52}_{-0.43}$ &
$1.63^{+0.44}_{-0.43}$ \\
		Fe \Ka /Si \Hea &
$2.17^{+0.54}_{-0.51}$ &
$1.08\pm{0.02}$ &
$0.94\pm{0.02}$ &
$1.17\pm{0.06}$ &
$0.99\pm{0.05}$ &
$0.54\pm{0.06}$ &
$11.4\pm{0.9}$ &
-- \\ \hline
	\end{tabular}
\begin{tablenotes}
\item[]  The errors are at 90\% C.L.
\item[a] Here, we used the spectrum extracted only from regions inside the SNR for category 0.
\item[b] Photon flux at 1 keV in units of $\rm 10^{-6}\ ph\ cm^{-2}\ s^{-1}$.
\end{tablenotes}
\end{threeparttable}
\end{table*}

The medium- and high-energy parts of the spectra in \figr\ref{fig:GMM_Tycho_spec} have clear line emission features. To investigate the spectral properties quantitatively, we fitted the spectra with a model of an absorbed power law for the continuum emission plus Gaussians for the emission lines of \Hea, \Lya, \Heb, \Heg, and \Lyb\ of Si, S, Ar, and Ca (excluding Ca~\Lyb ) and Fe~\Ka . The line centroids, widths, and intensities of weak lines such as Lyman lines and \Heg\ were tied to other strong lines following \citet{Hayato2010}. We assumed a hydrogen column density of $7\times 10^{21}\ \rm cm^{-2}$ \citep{CassamChenai2007} and standard ISM abundances \citep{Wilms2000}. The results of the model fitting are shown in \tabr\ref{tab:GMM_spec_paras} and \figr\ref{fig:GMM_Tycho_line_paras}.
In category~7, continuum emission is dominant; thus, the Fe \Ka\ line cannot be detected.

In the top and middle panels of \figr\ref{fig:GMM_Tycho_line_paras}, the centroid energy of Fe \Ka\ is higher, and the \Lya/\Hea, \Heb/\Hea\ line flux ratios of Si and S are higher, except for category~6.
These trends correspond spatially to higher ionization and temperature in the outer side of the SNR.
Especially in the NW region of the SNR, in which categories~1--5 form a layered structure, the physical parameters of ionization and the temperature appear to change the from inner to outer the SNR.
We analyse the NW region in detail in \secr\ref{sec:NW_ejecta}.
On the other hand, category~6, which is located on the rim in the eastern region of the SNR, has different characteristics.
For category~6, the centroid energy of Fe \Ka\ is highest among all the categories, but the \Lya/\Hea, \Heb/\Hea\ line flux ratios of Si and S are lower than the others.
We analyse the Fe knot region, including category~6, in detail in \secr\ref{sec:Fe_knot}.

In the bottom panel of \figr\ref{fig:GMM_Tycho_line_paras}, the line flux ratios of \Hea\ of S, Ar, or Ca to Si increase gradually from the inner side toward the outer side of the SNR.
On the other hand, the Fe \Ka /Si \Hea\ line flux ratios show a different trend.
These ratios decrease toward the outer side of the main SNR shell and are lowest in the region of category~5,
although the ratio for category~6, which is located the outer edge of the SNR, is only an order of magnitude higher than the others.
Category~6 clearly has different spectral features from the other regions.

Although it is difficult to fully understand how each feature in the raw data (i.e., spectral structure in this case) affects the latent expressions, we think the low-energy sides of the spectra also contribute significantly to the clustering because the low-energy part is divided into finer bins than the high-energy part is. For example, whereas the Fe~\Ka\ blend consists of only 1 energy bin in our binning method, the energy band of the Fe~L blend between 0.75 and 1.31~keV is divided into 10~bins.

The \ion{Fe}{xvii} line ($n=3\rightarrow2$) structure at $\sim$0.83~keV is divided into 4 bins and appears strongly in the spectra of categories~1--4. In addition, the Mg~\Ka\ line at $\sim$1.35~keV appears strongly only in categories~1 and 2, and the O~K line structure at $\sim$0.65~keV appears strongly in the spectra of categories~0, 1, and 7. We think that these structures contribute to distinguishing these categories from others.

\section{Detailed Analysis of the Regions Suggested by Machine Learning}\label{sec:analysis}

As shown in the previous section, the unsupervised machine learning method can discover spatial structures.
In this section, we choose two regions of the revealed structure and analyse them in detail.

\subsection{Spectral Analysis of the Fe Knot}\label{sec:Fe_knot}

\begin{figure}
\centering
\includegraphics[width=\columnwidth,keepaspectratio]{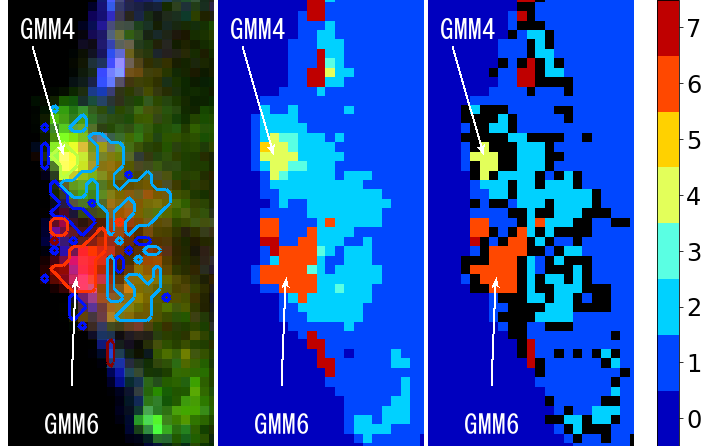}
\caption{Substructure of the Fe knot on the eastern rim of \tycho.
The {\it left panel} shows a three-colour image (red: Fe L blend band, 0.7--0.95 keV; green: Si~\Hea\ band, 1.75--1.95 keV; blue: continuum band: 4.5--5~keV).
In the {\it right panel}, for each spatial bin of the merged data from 2009, categories with the highest responsibility are assigned and colour-coded.
In the {\it right panel}, only spatial bins with a category having a responsibility above 90\% are selected and assigned the colours representing the categories. The other spatial bins, which have responsibilities below 90\%, appear black.
\protect\\
(A colour version of this figure is available in the online journal.)}
\label{fig:SE_Tycho_region_img}
\end{figure}

\begin{figure}
\centering
\includegraphics[width=\columnwidth,keepaspectratio]{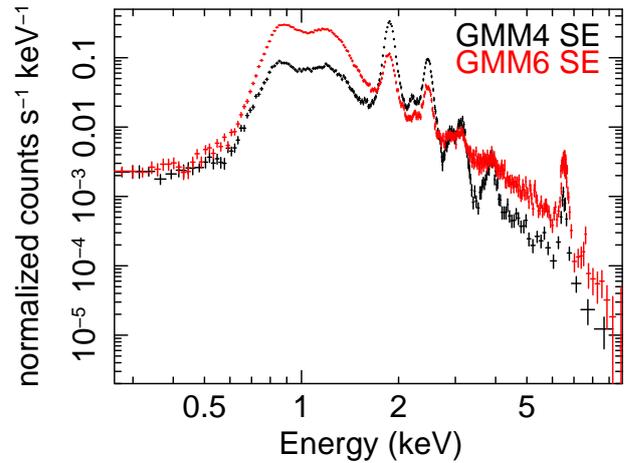}
\caption{Spectra of GMM categories 4 and 6 with GMM responsibilities above 90\% in the Fe knot. The black spectrum is GMM category 4, and the red one is category 6.}
\label{fig:SE_Tycho_spec}
\end{figure}

\begin{table*}
	\centering
 \caption{Best-fitting parameters for the spectrum of each GMM category in the Fe knot with responsibility $\ge$ 0.9}
	\label{tab:GMM_spec_paras_SE}
\begin{threeparttable}
	\begin{tabular}{cccccc}
		\hline \hline
		 & \multicolumn{5}{c}{Region} \\
		 & 1 & 2 & 4 & 6 & 7 \\ \hline
		  \multicolumn{6}{c}{Power Law} \\ \hline
		 Photon Index &
$2.97^{+0.13}_{-0.14}$ &
$2.51^{+0.27}_{-0.26}$ &
$1.92^{+0.60}_{-0.56}$ &
$2.85^{+0.14}_{-0.16}$ &
$2.66^{+0.15}_{-0.21}$ \\
		 Surface Brightness \tnote{a}&
$141^{+27}_{-25}$ &
$92^{+42}_{-28}$ &
$94^{+99}_{-38}$ &
$303^{+70}_{-64}$ &
$413^{+99}_{-120}$   \\ \hline
		  \multicolumn{6}{c}{Fe K$\alpha$ Lines} \\ \hline
		 Centroid (keV) &
$6.504^{+0.024}_{-0.021}$ &
$6.509^{+0.009}_{-0.008}$ &
$6.501\pm{0.018}$ &
$6.526^{+0.009}_{-0.010}$ &
 -- \\
 		Width (keV) &
$0.043^{+0.039}_{-0.047}$ &
$0.076\pm{0.012}$ &
$0.043^{+0.030}_{-0.056}$ &
$0.060^{+0.014}_{-0.016}$ &
 -- \\
		Surface Brightness \tnote{a}&
$0.86\pm{0.16}$ &
$2.66\pm{0.17}$ &
$5.19^{+0.80}_{-0.79}$ &
$5.43\pm{0.39}$ &
 --  \\ \hline
		  \multicolumn{6}{c}{K$\alpha$ Line Flux Ratio (\%)} \\ \hline
		 S \Hea /Si \Hea &
$37.3\pm{1.9}$ &
$31.3\pm{0.5}$ &
$34.7\pm{0.84}$ &
$37.2\pm{2.2}$ &
$27.9^{+6.5}_{-11}$ \\
		 Ar \Hea /Si \Hea &
$3.44^{+0.72}_{-0.67}$ &
$2.97^{+0.20}_{-0.21}$ &
$3.60^{+0.27}_{-0.28}$ &
$4.63^{+1.10}_{-0.95}$ &
$6.0^{+2.4}_{-2.3}$  \\
		 Ca \Hea /Si \Hea &
$1.06^{+0.39}_{-0.32}$ &
$1.02\pm{0.08}$ &
$1.03\pm{0.12}$ &
$1.28^{+0.52}_{-0.43}$ &
$2.1\pm1.5$  \\
		 Fe \Ka /Si \Hea &
$2.79\pm{0.51}$ &
$1.77\pm{0.11}$ &
$0.87\pm{0.13}$ &
$11.4\pm{0.9}$ &
 -- \\ \hline
	\end{tabular}
\begin{tablenotes}
\item[] The errors are at 90\% C.L.
\item[a] Surface brightness at 1 keV in units of $\rm 10^{-5}\ ph\ cm^{-2}\ s^{-1}\ arcsec^{-1}$.
\end{tablenotes}
\end{threeparttable}
\end{table*}

The Fe knot located along the eastern rim of \tycho\ represents unusual morphological features in which several iron-rich clumps outrun the FS.
The Fe knot can be divided into substructures, and \citet{Yamaguchi2017} analysed in detail these fine regions.
The regions defined by \citet{Yamaguchi2017} have the following counterparts in our analysis: `A' and `E', category 6; `B', category 4; `C' and `D', category 2; `X' and `Y', category 7.

We extracted spectra from the substructure representative of each category with responsibilities above 90\%.
The results of fitting by the model described in the previous section are summarised in \tabr\ref{tab:GMM_spec_paras_SE}.

Figure \ref{fig:SE_Tycho_spec} shows the spectra of the regions of categories~4 and 6 in the Fe knot, which correspond to the regions marked `GMM4' and `GMM6' in \figr\ref{fig:SE_Tycho_region_img}, respectively, with responsibilities above 90\%.
The spectrum of category~4 has enhanced IME line radiation, although one of the category~6 features has weaker IME lines and stronger Fe~\Ka\ emission.
The regions of the Fe knot corresponding to categories 4 and 6 are characterised by the lowest and highest ratios of the Fe~\Ka\ and Si~\Hea\ line fluxes, respectively.
The Fe/Si flux ratio in category 6 is also quite high, as shown in the bottom panel of \figr\ref{fig:GMM_Tycho_line_paras}; thus, the region has a spectral feature unique to not only the entire SNR but also the Fe knot.
The machine learning extracted a characteristic structure in which the Fe emission is much stronger.

The \Hea\ line flux ratios of S, Ar, or Ca to Si in categories 1, 2, 4, and 6 are approximately constant in the Fe knot.
This implies a constant electron temperature in the knot, in agreement with \citet{Yamaguchi2017}.

The S~\Hea /Si~\Hea\ line flux ratios in categories 1 and 2 in the Fe knot are higher than those in the entire SNR (\tabr\ref{tab:GMM_spec_paras}).
This reflects a higher ionization state in the Fe knot, which is located on the SNR rim (i.e., at a large radius), than that typical of the shocked and unshocked ejecta in the inner part of the SNR, where categories 1 and 2 are most common.

The Fe~\Ka /Si~\Hea\ line flux ratios in categories 1 and 2 in the Fe knot are higher than those in the entire SNR (\tabr\ref{tab:GMM_spec_paras}).
However, this ratio for category 4 in the Fe knot is comparable to that in the entire SNR.
As shown in the left image of \figr\ref{fig:SE_Tycho_region_img}, the clump emitting Fe coincides with the blob emitting IME in the category 4 region.

\subsection{Spectral Analysis of NW Ejecta}\label{sec:NW_ejecta}

\begin{figure}
\centering
\includegraphics[width=8cm,keepaspectratio]{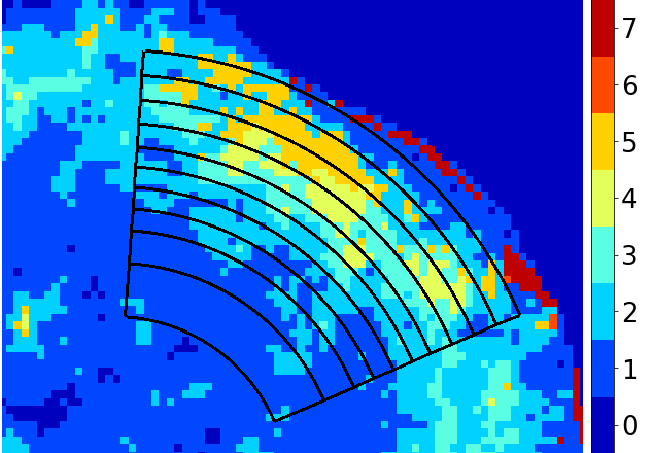}
\caption{Annular regions in the NW region of \tycho.
\protect\\
(A colour version of this figure is available in the online journal.)}
\label{fig:NW_Tycho_region_img}
\end{figure}

\begin{figure}
\centering
\includegraphics[width=\columnwidth,keepaspectratio]{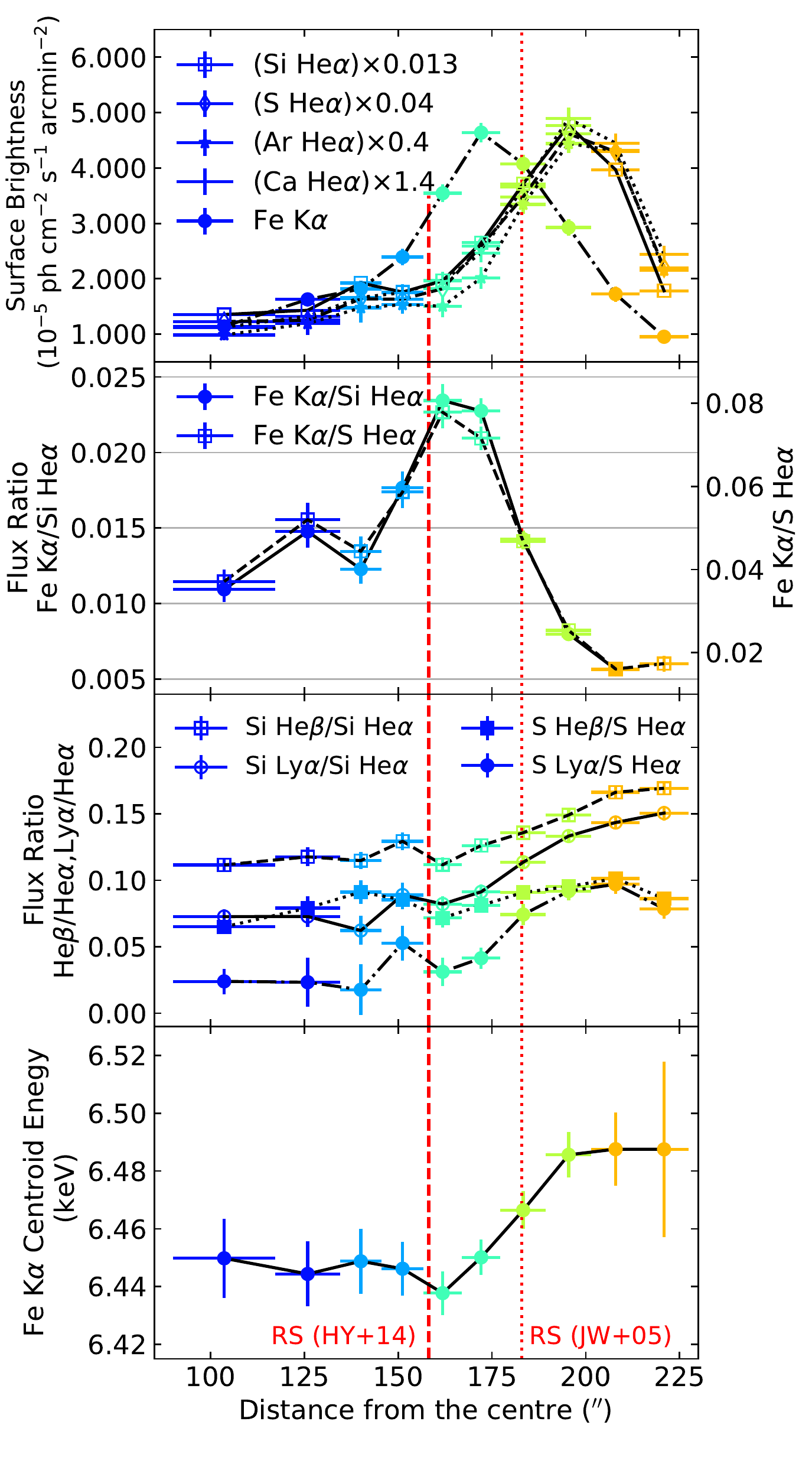}
\caption{Radial dependence of the best-fitting Gaussian parameters in the NW annular regions. The {\it top panel} shows the surface brightness of the lines of Si~\Hea, S~\Hea, Ar~\Hea, Ca~\Hea, and Fe~\Ka,
the {\it upper middle panel} shows the flux ratios of Fe~\Ka/Si~\Hea\ or Fe~\Ka/S~\Hea,
the {\it lower middle panel} shows the flux ratios of \Heb/\Hea\ and \Lya/\Hea\ of Si and S,
and the {\it bottom panel} shows the centroid energies of Fe~\Ka\ lines. The errors are at 90\% C.L. The RS positions determined by \citet{Yamaguchi2014} and \citet{Warren2005} are indicated by the vertical dashed line and the dotted line, respectively.}
\label{fig:NW_Tycho_line_plots}
\end{figure}

In the NW region of \tycho, categories~1--5 and 7 are layered.
We divided the NW part of the SNR into annular regions along the layered features representing the categories, as shown in \figr\ref{fig:NW_Tycho_region_img}.
The centre of the annuli was determined to be R.A. = 00$^{\rm h}$25$^{\rm m}$17\fs754, +64\degr08\arcmin06\farcs549 so that those annuli align with the layered structure.
The regions were labelled NW1 to NW10 from the inner to the outer side.
We extracted the spectra from these regions, adopting a background spectrum extracted from a box region outside of the SNR. We performed model fitting in the 1.7--9~keV band using the same models as in Section~\ref{sec:spec_ana_GMM_All_tycho} to investigate the line emission of IMEs and Fe.

The surface brightness peak of the Fe~\Ka\ line forms the innermost layer, as compared to the \Hea\ lines of Si, S, Ar, and Ca, in the top panel of \figr\ref{fig:NW_Tycho_line_plots}.
This trend was previously seen in {\it ASCA} observations \citep{Hwang1997} and {\it XMM-Newton} observations \citep{Decourchelle2001}.

The upper middle panel of \figr\ref{fig:NW_Tycho_line_plots} shows the Fe~\Ka /Si~\Hea\ or Fe~\Ka /S~\Hea\ flux ratios, which peak between regions NW5 and NW6 (156.7--177.3~arcsec).
They reflect the difference in the peak positions of the Fe~\Ka\ and \Hea\ of Si or S line emission.

In the lower middle panel of \figr\ref{fig:NW_Tycho_line_plots},
the line flux ratios of \Lya /\Hea\ and \Heb /\Hea\ of Si and S increase from the inner region NW5 to the outer one NW10 (156.7--227.3~arcsec).
These trends correspond to a radial gradient of temperature or ionization, which is higher on the outer side of the SNR.
A temperature gradient in the ejecta is suggested, as it is hottest near the RS and cooler near the CD,
because the Fe~\Ka\ emission peak is located interior to those of Fe~L and Si \citep{Hwang1997,Decourchelle2001}.
Furthermore, \citet{Yamaguchi2014} showed electron heating near the RS of \tycho.
If we attribute our results to a temperature gradient, our findings imply an opposite trend to that inferred in these works.
On the other hand, a variation in the Fe ionization state near the RS \citep{Yamaguchi2014} was reported.
Moreover, a radial gradient of the ionization age was suggested by the Si~\Hea /S~\Hea\ flux ratio \citep{Lu2015} and Fe~\Ka\ centroids \citep{Sato2017}.
Thus, our results are consistent with these works if we interpret the radial dependence of the line flux as arising from the ionization age gradient induced by RS propagation.

Then we fit the spectra with models in the 4.2--10~keV band to investigate the \Ka\ and \Kb\ lines of Fe and the \Ka\ lines of secondary Fe-peak elements (Cr and Mn).
The Gaussian widths of the Fe~\Kb, Cr~\Ka, and Mn~\Ka\ lines are linked to those of Fe~\Ka.

The centroid energies of the Fe~\Ka\ lines shown in the bottom panel of \figr\ref{fig:NW_Tycho_line_plots} are flat between NW1 and NW5 (90.0--167.0~arcsec) and begin to increase at NW5 around 162~arcsec.
By contrast, the line width of Fe~\Ka\ (except in the outermost region, NW10) and the centroid energies of Fe~\Kb\ do not change significantly.
The peak of the surface brightness of Fe~\Ka\ is in the NW6 region between 177.3 and 189.4~arcsec, and the peak of Fe~\Kb\ is in the inner region NW5 between 156.7 and 167.0~arcsec, which is consistent with the results of \citet{Yamaguchi2014}.

\citet{Warren2005} estimated the averaged RS radius to be 183~arcsec. The RS radius is located in region NW7, which coincides with the turning point of the surface brightness of IME~\Hea\ and Fe~\Ka\ in the top panel of \figr\ref{fig:NW_Tycho_line_plots}.
By contrast, \citet{Yamaguchi2014} estimated the RS radius as 158~arcsec in the NW quadrant, which is interior to and more realistic than the former one.
The RS radius is located near the boundary of regions NW4 and NW5.
It coincides with the turning point of the centroid of Fe~\Ka\ in the bottom panel of \figr\ref{fig:NW_Tycho_line_plots}.
Thus, it seems that categories~3--5 are ejecta shocked by the RS, and most of categories~1 and 2 are located inside the RS in projection.

In \figr\ref{fig:NW_Tycho_line_plots}, the transitions of the flux ratios of line emission or the line centroid energy described above appear at the RS position.
Thus, the coincidences suggest that the features reflect the plasma state of the ejecta caused by RS heating.

We also note the detection of Fe~\Kb\ in regions NW3--NW8 excluding NW7, Cr~\Ka\ in regions NW6 and NW8, and Mn~\Ka\ in region NW8 with $3\sigma$ or more.

\section{Discussion and Conclusions}\label{sec:discussion}

We implemented an unsupervised machine learning method combining the VAE and GMM, where the dimensions of the observed data are reduced by the VAE, and clustering in feature space is done by the GMM, and applied the method to Tycho's SNR, one of the best-known SNRs.

Our unsupervised machine learning method automatically revealed spatial structures which have been discussed in the literature \citep[see, e.g.,][]{Yamaguchi2017}.
This demonstration shows that the method is a powerful tool for data analyses that makes it possible to exploit the rich information contained in data obtained by X-ray observations of SNRs.
It may be possible to discover SNR physics by post-training analyses using the results of machine learning.

It is also worth noting that the method discovered the spatial structures automatically, although no spatial information was used in the model.
This means that the method can extract physical feature based only on the spectral information.

As demonstrated in Sections~\ref{sec:demo} and \ref{sec:analysis}, the VAE extracts features using the relative intensities of lines as well as the properties of the continuum spectrum.
These characteristics of thermal X-ray spectra reflect the plasma conditions (e.g., temperature, ionization, elemental abundances, and electron or ion densities).
When the data distribution in feature space is categorized by the GMM, the entire region is divided into a small number of clusters. As shown by our analysis, clustering can reveal both sharp, knot-like features and continuous changes in physical parameters.
Sharp structures are classified as a single category. For example, the Fe-rich blob in the Fe knot on the eastern rim of the SNR, shown in \secr\ref{sec:Fe_knot}, is assigned to category~6.
By contrast, if physical parameters change gradually, clustering may result in a layered spatial structure like that seen in the NW regions of the SNR (see \secr\ref{sec:NW_ejecta} for details).

The reason that each individual spectrum is classified in a certain category is not yet clear from the network outputs but needs to be investigated and interpreted by human experts, as we mentioned in \secr\ref{sec:spec_ana_GMM_All_tycho}. It would be useful if the network itself provided the reason, e.g., by highlighting the spectral features that cause the spectrum to be assigned to a particular category. Unveiling the reasoning process of the network is a significant problem and is beyond the scope of the current work \citep[see, e.g.,][for recent reviews]{Smilkov2017}.

Developing such methods is important for making the best use of the currently available data and for addressing the growing quality and quantity of future observational data.
Model fitting of spectra is generally time-intensive; thus, the difficulty of spectral analysis is expected to increase steeply with successful implementation of an X-ray microcalorimeter \citep[such as {\it Athena};][]{AthenaX-IFU2018SPIE}. The suggested unsupervised method can reveal characteristic features directly from raw observational data without spectral model fitting.
It can be an efficient tool to define regions for spectral extraction.

Our method implemented in this work is not limited to SNRs and can be applied to other classes of sources such as galaxy clusters.
The method is equally applicable to temporally and spatially variable data, because the training uses only spectral information.
Furthermore, our method can also be applied to other energy bands; e.g., it is expected to have good applicability to radio observational data, which contain spatial, temporal, and velocity information (i.e., they have the same dimensions as X-ray data: spatial, temporal, and spectral information).

The deep learning architecture can be improved.
In this method, it is a problem that the VAE used to reduce the dimensions of the data tends to form a single peak distribution around 0 in feature space; thus, the boundaries of the extracted data distribution are not clear.
Using the architecture of the Wasserstein autoencoder \citep[WAE;][]{Tolstikhin2017arXiv} or Gaussian Mixture VAE \citep[GMMVAE;][]{Dilokthanakul2016arXiv} may improve the structure of the latent manifold.
A model using convolutional layers, e.g., a convolutional VAE, can be applied to use the spatial information in a data set.

\section*{Acknowledgements}


We thank Dr Hiroya Yamaguchi, Masato Taki, and Dmitry Khangulyan for useful information and discussion.
This work was partially supported by the Japan Society for the Promotion of Science (JSPS) KAKENHI Grants (Number 18H03722 and 18H05463).
This work was supported by the Astro-AI working group in the RIKEN iTHEMS.




\bibliographystyle{mnras}

\bibliography{bib_vae_tycho}








\bsp	
\label{lastpage}
\end{document}